\documentclass[a4paper]{jpconf}
\usepackage{graphicx}
\begin{document}
\newcommand{\mc}{\multicolumn}
\newcommand{\bce}{\begin{center}}
\newcommand{\ece}{\end{center}}
\newcommand{\be}{\begin{equation}}
\newcommand{\ee}{\end{equation}}
\newcommand{\bea}{\vspace{0.25cm}\begin{eqnarray}}
\newcommand{\eea}{\end{eqnarray}}
\def\NCA{{Nuovo Cimento } A }
\def\NIM{{Nucl. Instrum. Methods}}
\def\NPA{{Nucl. Phys.} A }
\def\PLA{{Phys. Lett.}  A }
\def\PRL{{Phys. Rev. Lett.} }
\def\PRA{{Phys. Rev.} A }
\def\PRC{{Phys. Rev.} C }
\def\PRD{{Phys. Rev.} D }
\def\ZPC{{Z. Phys.} C }
\def\ZPA{{Z. Phys.} A }
\def\PTP{{Progr. Th. Phys. }}
\def\LNC{{Lett. al Nuovo Cimento} }

\title{Experimental tests of hidden variable theories from dBB to Stochastic Electrodynamics}

\author{Marco Genovese $^1$, Giorgio Brida $^1$, Marco Gramegna$^1$, Fabrizio Piacentini$^1$, Enrico Predazzi$^2$,
Ivano Ruo-Berchera$^1$}
\address{$^1$I.N.RI.M., Str.
delle Cacce 91, I-10135 Torino, Italy \\ $^2$ Dip. Fisica Teorica
Univ. Torino and INFN, via P. Giuria 1, 10125 Torino, Italy}

\ead{genovese@inrim.it}

\begin{abstract}
The studies concerning the possible existence of a deterministic
theory, of which quantum mechanics would be an approximation, date
to the celebrated 1935 Einstein-Podolsky-Rosen paper. Since Bell's
proposal of 1964 various experiments were addressed to a general
experimental test of local hidden variable theories, leading to
strong indications favourable to Standard Quantum Mechanics.
Nevertheless, detection loophole still persists. In this, after a
short presentation of recent PDC photon experiments, we will present
our recent works in this field and in particular a conclusive
negative test of stochastic electrodynamics. Finally, we will also
mention  possible tests of non-local deterministic models and give
some detail on our test of the dBB model.

\end{abstract}

\section{Introduction}

The quest for a realistic theory of which quantum mechanics would be
an approximation dates to more than 70
 years and in the last years has accelerated due to both experimental and theoretical progresses \cite{prep}.

Already in 1935 Einstein-Podolsky-Rosen \cite{EPR}, analysing the
measurement on an
 entangled state, proposed that Quantum Mechanics
(QM) could be an incomplete theory, representing a statistical
approximation of a complete deterministic theory: this was the birth
of Local Hidden Variable Theories (LHVT), where the values of the
observables are fixed by some hidden variable and probabilistic
predictions become epistemic, being due to our ignorance of the
hidden variables.

A subsequent fundamental progress in discussing Local Hidden
Variable  was Bell's discovery that any theory of this kind must
satisfy certain inequalities that can be violated in QM leading in
principle to a possible  experimental test of the validity of the
standard interpretation of QM (SQM)  compared to LHVT.

Since then, many interesting experiments have been devoted to  test
Bell inequalities, the most interesting of them using photon pairs
\cite{Mandel,asp,franson,type1,type2}, leading to a substantial
agreement with quantum mechanics and disfavouring LHVT. Up to now,
however, no experiment has yet been able to exclude definitively
such theories. In fact, due to the low total detection efficiency,
so far one has always been forced to introduce at least one further
additional hypothesis \cite{santos},  i.e. the observed sample of
particle pairs is a faithful subsample of the initial set of pairs.
This problem is known as { \it detection or efficiency loophole}.
The research for
 new experimental configurations able to overcome the detection loophole is, of course, of the utmost interest.

In the 90's  a relevant progress in the direction of eliminating
this loophole has been obtained by using Parametric Down Conversion
(PDC) process.

This technique \cite {Mandel} has been largely employed to produce
"entangled" photon pairs, i.e. pairs of photons described by a
common wave function that cannot be factorized into the product of
the wave functions of individual photons.

The generation of entangled states by parametric down conversion
(PDC) has replaced other techniques, such as the radiative decay of
excited atomic states, as it was in the celebrated experiment of A.
Aspect et al. \cite{asp}, for it overcomes some previous limitation.
In particular, it overcomes the poor angular correlation of atomic
cascade photons, that is at the origin of the small total efficiency
of this type of experiments where one is forced to select a small
subsample of the produced photons, leading inevitably to  the
detection loophole, since PDC presents angular correlations better
than 1 mrad

The first experiments using this technique were performed with
type I PDC, which gives phase and momentum entanglement and can be
used for a test of Bell inequalities using two spatially separated
interferometers \cite{franson}, as realized by Ref.\cite{type1}.
The use of beam splitters, however, strongly reduces the total
quantum efficiency.

In alternative, a polarization entangled state can be generated
\cite{ou}. However, in the earlier attemps, in generating this
state, half of the initial photon flux was lost  and the efficiency
loophole could not be eliminated even in principle \cite{santos}.

More recently experiments where a polarization entangled state is
directly generated, have been realized using Type II PDC
\cite{type2,prep}. This scheme has allowed a much higher total
efficiency than the previous ones at the price of delicate
compensations for having identical arrival time of the ordinary and
extraordinary photons. Such an efficiency, however, is  at most
$\approx 30 \%$, still far from the value of $0.81$ required for
eliminating the detection loophole for a maximally entangled state.
In addition, experiments studying equalities among correlations
functions rather than Bell inequalities \cite{GHZ,dem} are very far
from giving a loophole free test of local realism \cite{garuccio}. A
large interest remains therefore for new experiments increasing
total quantum efficiency in order to reduce and finally overcome the
efficiency loophole \footnote{It must be acknowledged that a recent
experiment \cite{Win} based on the use of Be ions has reached very
high detection efficiencies (around 98 \%), largely sufficient for
closing the detection loophole. However, in this case not only space
like separation required for closing locality loophole was not
satisfied, but the two subsystems (the two ions) were even not
really separated during the measurement. Therefore, this experiment
cannot be considered a real implementation of a loophole free test
of Bell inequalities, even if it represents a relevant progress in
this sense. Also tests with pseudoscalar mesons, albeit interesting,
look far from a suitable way for eliminating detection loophole
\cite{k}.}.

On the other hand, even if  conclusive loophole free experiments
will allow to exclude LHVT beyond any possible doubt, room will
remain for non local Hidden Variable Theories \cite{prep,adler} or
models were "true" degrees of freedom, on which Bell inequalities
tests should be performed, are at large (Plank) scales
\cite{thooft}.

In this paper we will present some  experimental investigation in
this field  based on the production of various polarization
entangled states in  both cases: type I and type II sources
\footnote{Incidentally, it must be noticed that the perfect fitting
of SQM predictions to data obtained by varying one polarizer angle
for various different settings of the other one both obtained in our
experiments as in many other ones \cite{prep,shap} conclusively
falsifies LHVT based on the  violation of rotational invariance,
e.g. \cite{min}.}. In particular, we will describe  a bright source
of (non-maximally) polarization entangled states of photons and its
use to test Bell inequalities. Then, we will describe its
application to exclude a specific local realistic theory (stochastic
electrodynamics), that survived previous experiments due to
detection loophole. Finally, we will mention that a modification of
this apparatus has been used for a first test of de Broglie-Bohm
theory.

\section{Bright source of non-maximally polarization entangled photons}

With the purpose of building a bright source of (non-maximally)
entangled states, we have considered \cite{napoli} the possibility
of generating a polarization entangled state via the superposition
of the spontaneous fluorescence emitted by two non-linear crystals
(rotated in order to have orthogonal polarization) driven by the
same pumping laser \cite{hardy}. The crystals are in cascade along
the propagation direction of the pumping laser and the superposition
is obtained by using an appropriate optics. If the path between the
two crystals is smaller than the coherence length of the laser, the
two photon paths are indistinguishable and a polarization entangled
state is created. In fact, applying the evolution operator given by
the PDC Hamiltonian one has, in first order of the perturbation
expansion (a good approximation in low gain regime): \be \vert \Psi
\rangle =\vert vacuum \rangle + f_1 V_1 \vert H \rangle \vert H
\rangle + f_2 V_2 \vert V \rangle \vert V \rangle \label{PsiH} \ee
where $f_i$ takes into account the properties of crystal $i$
($|f_i|^2$ is the fraction of incident light down converted by the
non-linear crystal) and $V_i$ the pump intensity at the crystal $i$.

The possibility of obtaining easily a non maximally entangled state
(where $V_1 f_1$ and $V_2 f_2$ are different) is very interesting,
since it has been shown that for non maximally entangled pairs the
lower limit on the total detection efficiency for eliminating the
detection loophole is reduced to 0.67 \cite{eb} (compared with 0.81
for maximally entangled states). However, it must be noticed that,
for non-maximally entangled states, the largest discrepancy between
quantum mechanics and local hidden variable theories is also
reduced: thus a compromise between a lower total efficiency and a
still sufficiently large value of this difference is necessary when
realizing an experiment addressed to overcome the detection
loophole.

In our experimental set up two $LiIO_3$ crystals (10x10x10 mm,
$d_{31} = 3.5 \pm 0.4$ pm/V \cite{BGN}) were placed along the pump
laser propagation, 250 mm apart,  a distance smaller than the
coherence length of the pumping laser. This guaranteed
indistinguishability in the creation of a couple of photons in the
first or in the second crystal. A couple of plano-convex lenses of
120 mm focal length centered in between, focalized the spontaneous
emission from the first crystal into the second one maintaining  the
angular spread. A hole of 4 mm diameter was drilled into the center
of the lenses to allow transmission of the pump radiation without
absorption and, even more important, without adding stray-light,
because of fluorescence and diffusion of the UV radiation.   A small
quartz plate (5 x5 x5 mm) in front of the first lens of the
condensers compensated pump birefringence. Finally, a
half-wavelength plate immediately after the condenser rotated the
polarization of the laser beam and excited in the second crystal a
spontaneous emission cross-polarized with respect to the first one.
Two correlated emissions at 633 and 789 nm were selected and, after
the polarizers, detected by two avalanche photodiodes with active
quenching. The output signals from the detectors were routed to a
two channel counter, in order to have the number of events on single
channel, and to a Time to Amplitude Converter circuit, followed by a
single channel analyzer, for selecting and counting coincidence
events.

A very interesting degree of freedom of this configuration is given
by the fact that by tuning the pump intensity between the two
crystals, one can easily select the value of $f=(f_2 V_2)/(f_1
V_1)$, that determines how far from a maximally entangled state
($f=1$) the produced state is. This is a fundamental property, which
permits to select the most appropriate state for the experiment

Our main result was the observed violation of the Clauser-Horne (CH)
inequality

\begin{equation}
CH=N(\theta _{1},\theta _{2})-N(\theta _{1},\theta _{2}^{\prime
})+N(\theta _{1}^{\prime },\theta _{2})+N(\theta _{1}^{\prime
},\theta _{2}^{\prime })-N(\theta _{1}^{\prime },\infty )-N(\infty
,\theta _{2}) < 0 \label{eq:CH}
\end{equation}
valid for every local realistic theory. In (\ref{eq:CH}),
$N(\theta _{1},\theta _{2})$ is the number of coincidences between
channels 1 and 2 when the two polarizers are rotated to an angle
$\theta _{1}$ and $\theta _{2}$ respectively ($\infty $ denotes
the absence of selection of polarization for that channel).

In our case we have generated a state with $f \simeq 0.4$: in this
case the largest violation of the inequality is reached for
$\theta_1 =72^o.24$ , $\theta_2=45^o$, $\theta_1 ^{\prime}=
17^o.76$ and $\theta_2 ^{\prime}= 0^o$. Our experimental result is
$CH = 513 \pm 25$  \cite{n1,n2}.

This clear violation of the CH inequality with non maximally
entangled states represented an interesting progress toward a
loophole free test of local realism \footnote{In Ref.\cite{k2} a
test of local realism was also performed with non-maximally
entangled states by using equalities \cite{HardyE}.}. Furthermore,
this experiment \cite{n2} allowed a clear negative test of
stochastic electrodynamics \cite{sed}, a theory built for
reproducing quantum electrodynamics results in a classical field
theory framework when a zero-point field is introduced. In its
subpart concerning the quantum properties of radiation, named
stochastic optics, it was forecasted that Bell inequalities should
not be violated below a certain level of detection rate \cite{cas}.
Indeed a clear violation of CH inequality was observed in our
experiment even  well  below this threshold (by many orders of
magnitude).

A little more in detail, no violation of Bell inequalities should
be measured when the single detection rate $R_S $ is lower than

\begin{equation}
R_S < { \eta F^2 R_c^2 \over 2 L d^2 \lambda \sqrt{ \tau T} }
\label{rate}
\end{equation}
where $\eta$ is the detection quantum efficiency, F is the focal
distance of the lens in front of detectors, $R_c$ is the radius of
the active area of the non-linear medium where entangled photons are
generated, $\tau$ is the coherence time of incident photons, d is
the distance between the non-linear medium and the photo-detectors,
$\lambda$ is the average wavelength of the detected photons. L and T
are two free parameters which are less well determined by the
theory: L can be interpreted as the active depth of the detector,
while T is the time needed for the photon to be absorbed, which
should be  approximately less than 10 ns, i.e. in a first
approximation, the length of the wave packet divided for the
velocity of light.

By introducing the parameters of our experiments in Eq. \ref{rate}
one obtains $ T > 1 s$, which is largely above any realistic
estimate.

 Besides, as a further
test of stochastic optics we also searched for a SPontaneous Up
Conversion (SPUC) emission predicted in this theory \cite{puc} with
an intensity comparable with SPDC. More in details, following the
indications of \cite{spuc}, we pumped with both a diode laser at 789
nm (50 mW power) and a Neodimium-Yag laser beam (1064 nm wave
length, 0.51 W power) a 1.5 cm Lithium Iodate crystal in the
configuration were a stimulated emission was emitted when a UV pump
(351 nm Argon laser beam, 0.3 W) was present. In the same
configuration SPUC was expected when the UV beam was turned off. We
did not observe any emission by monitoring the emission after the
crystal with a  ccd camera \cite{nosW}, being the SPUC signal at
least 160 times smaller than the PDC one.

Even a more severe test was later performed by  scanning
substantially all the possible angles for the emission when a 5 mm
BBO crystal was pumped by a 789 nm wave length, 90 mW power, diode
laser beam. For the sake of completeness the angle between pump
laser and crystal optical axis was also systematically varied of
small steps. Again no SPUC emission was observed up to two orders
of magnitude less than SPDC emission (after having normalized for
the different pump power).

Thus, taken together, all these negative results clearly falsify
this theory\footnote{Incidentally, also experiments at single photon
level where the parameter $\alpha$ (see \cite{gran} for a
definition) is susbstantially zero \cite{wpd} appear not to be
describable in such a theory.}.

Finally, we would like to mention that, by substituting in the two
crystals set-up for generating polarization entangled states the
second crystal with a double slit, it was possible to realize the
experiment proposed by Ref.\cite{parta} for testing standard quantum
mechanics against de Broglie-Bohm theory. In extreme synthesis, two
theoretical groups \cite{parta} proposed that when each of two
identical bosonic particles cross one of a double slit at the same
time, they never cross the symmetry axis of the slits at variance
with SQM predictions \footnote{For other proposals of possible
experimental tests of dBB against SQM see \cite{ds}}. Contrary to
this prediction we clearly observed coincidences of identical
photons (702 nm PDC conjugated photons) in the same semiplane
\cite{dBB} after crossing each a 10 micrometers slit (being the two
slits separated by 100 micrometers), e.g. with 35 acquisitions of
30' each we obtained an average of $78 \pm 10$ coincidences per 30'
when both the detectors were clearly in the same semiplane.

Even if the validity of this theoretical proposal is still under
discussion \cite{disc}, our result represents, as far as we know,
the first experimental attempt to test dBB theory. If this
theoretical prediction will be accepted as correct, then our
experiment could be interpreted as a conclusive test of dBB theory
(at least) in the version where photons have trajectories.

\section{Conclusions}

In this paper we have reviewed, after a general introduction to this
researches,  part of our work on testing realistic alternative to
standard quantum mechanics. In particular, a bright source of
non-maximally polarization entangled states and its applications to
moving toward a loophole free Bell inequalities experiment were
described. We also discussed how the analysis of the results
obtained  allows a falsification of stochastic electrodynamics.
Finally, we have hinted at the results obtained with a modification
of this set-up conceived to test SQM against dBB according to the
predictions of Ref.\cite{parta}.

\subsection{Acknowledgments}

We acknowledge   the support of MIUR (FIRB RBAU01L5AZ-002),
Fondazione San Paolo and Regione Piemonte (E14).
\section{References}

\subsection{Reference lists}

\end{document}